\documentclass[12pt]{iopart}
\usepackage{iopams}
\usepackage{graphicx}
\usepackage[T1]{fontenc}
\usepackage{setstack}

 \newcommand{\Rh}{\widehat{R}}

\newcommand{\av}[1]{\langle{#1}\rangle}

 \newcommand{\tdil}[2]{{\rm d} {#1} / {\rm d} {#2}}

 \newcommand{\df}{\ {\overset {\rm def} =}\ }

 \newcommand{\dril}[2]{{{\rm d} {#1}} / {{\rm d} {#2}}}
 \newcommand{\pdril}[2]{{\partial {#1}} / {\partial {#2}}}

\begin{document}

\review[Inhomogeneous cosmological models]{Inhomogeneous cosmological models:
exact solutions and their applications}


\author{Krzysztof Bolejko$^{1}$,
Marie-No\"elle C\'el\'erier$^{2}$,
Andrzej Krasi\'nski$^{3}$}

\address{$^{1}$ Astrophysics Department,
University of Oxford, Oxford OX1 3RH, UK}
\address{$^{2}$ Laboratoire Univers et Th\'eories,
Observatoire de Paris, CNRS, Universit\'e Paris Diderot, 5 place Jules Janssen,
92190 Meudon, France}
\address{$^{3}$ Nicolaus Copernicus Astronomical Center, Bartycka 18, 00-760
Warsaw, Poland}

\ead{Krzysztof.Bolejko@astro.ox.ac.uk,
marie-noelle.celerier@obspm.fr,
akr@camk.edu.pl}

\begin{abstract}
Recently, inhomogeneous generalisations of the Friedmann -- Lema\^{\i}tre --
Robertson -- Walker cosmological models have gained interest in the
astrophysical community and are more often employed to study cosmological
phenomena. However, in many papers the inhomogeneous cosmological models are
treated as an alternative to the FLRW models. In fact, they are not an
alternative, but an exact perturbation of the latter, and are gradually becoming
a necessity in modern cosmology. The assumption of homogeneity is just a first
approximation introduced to simplify equations. So far this assumption is
commonly believed to have worked well, but future and more precise observations
will not be properly analysed unless inhomogeneities are taken into account.
This paper reviews recent developments in the field and shows the importance of
an inhomogeneous framework in the analysis of cosmological observations.
\end{abstract}


\maketitle

\section{Introduction}

For this article, we define inhomogeneous cosmological models as follows: they
are those exact solutions of Einstein's equations that contain at least a
subclass of nonvacuum and nonstatic Friedmann -- Lema\^{\i}tre -- Robertson --
Walker (FLRW) solutions as a limit. The reason for this choice is that such FLRW
models are universally recognised as a good first approximation to a realistic
description of our actual Universe, so it makes sense to consider only those
other models that have a chance to be a still better approximation. Models that
do not include an FLRW limit would not easily fulfil this condition.

Among the models so defined we chose for a more detailed description only the
Lema\^{\i}tre \cite{Lema1933} -- Tolman \cite{Tolm1934} (LT) and Szekeres
\cite{Szek1975} models because they were the basis for the greatest number of
papers aimed at physical and astrophysical interpretation. The other
inhomogeneous models are only partly listed and some of them are briefly
described.

The LT and Szekeres models describe the evolution of the Universe in the
post-recombination era, in which only gravitational interactions play a role.
The matter source in them is dust, i.e. a perfect fluid with zero pressure (the
generalisation to nonzero cosmological constant is known, but less frequently
used). Thus, they should not be considered for application to pre-recombination
epochs, in which the pressure cannot be neglected. In particular, they should
not be applied to the inflationary epoch. Also, they are not suitable for
including ``dark energy'' in any other form than cosmological constant. On the
contrary, these models are sometimes used to explain observational results
attributed to the ``dark energy'' by effects of inhomogeneities in ordinary
matter. They are meant to be a replacement for the linearised perturbations of
the FLRW models and for methods describing backreactions with the help of
averaged quantities. Because of their symmetries (LT) and quasi-symmetries
(Szekeres) they apply to less general situations than the perturbative
calculations, but their advantage is that they fulfil the Einstein equations
exactly. Therefore, as long as we believe that general relativity is the correct
theory of gravitation, the inhomogeneous models can be extrapolated arbitrarily
far into the future and are not constrained by any ``regimes''.

The main body of this review is devoted to a description of those observed
effects that can be explained using the LT and Szekeres models. The penultimate
section deals with misuses, errors and misconceptions existing in the literature
on the inhomogeneous models.

\section{Exact solutions of Einstein's equations that can be applied as
inhomogeneous cosmological models}\label{ExactSolutions}

The total number of papers in which such solutions were derived or discussed was
approx. 750 until
1994 \cite{Kras1997}. No-one has updated that statistic. No generalisations of
models known until 1994 have been reported in later years. However, the
Lema\^{\i}tre \cite{Lema1933} -- Tolman \cite{Tolm1934} (LT) models have become
popular as a basis for astrophysical considerations, and the same is happening
recently with the Szekeres \cite{Szek1975} models. The current number may well
be over 1000. We begin by recalling the general classification scheme
\cite{Kras1997}.

\medskip

{\underline {(1) The Szekeres - Szafron (S--S) family \cite{Szek1975,Szaf1975}}}

\medskip

These models are invariantly defined by the following properties
\cite{SzCo1979}:

\noindent 1. They obey the Einstein equations with a perfect fluid source.

\noindent 2. The flow-lines of the perfect fluid are geodesic and nonrotating.

\noindent 3. The hypersurfaces orthogonal to the flow-lines are conformally
flat.

\noindent 4. The Ricci tensor of those hypersurfaces has two of its eigenvalues
equal.

\noindent 5. The shear tensor has two of its eigenvalues equal.

Because of property 2, in comoving coordinates the pressure depends only on
time. Thus the barotropic equation of state, the most popular one in the
astrophysics community, reduces the Szafron metric to the Friedmann --
Lema\^{\i}tre -- Robertson -- Walker (FLRW) class. The only nontrivial solutions
in the S--S family that can be reasonably applied in cosmology are the Szekeres
metrics \cite{Szek1975}, in which the source is dust (a perfect fluid with zero
pressure). This is a good model for the later phases of the evolution of the
Universe, in which gravitation plays a dominant role and large-scale
hydrodynamical processes have come to an end.

The metric of the Szekeres solutions is
\begin{equation}\label{e1}
{\rm d} s^2 = {\rm d} t^2 - {\rm e}^{2 \alpha(t, x, y, r)} {\rm d} r^2- {\rm
e}^{2 \beta(t, x, y, r)} \left({\rm d} x^2 + {\rm d} y^2\right).
\end{equation}
The coordinates of (\ref{e1}) are comoving so that $u^{\mu} = {\delta^{\mu}}_0$.
There are two families of Szekeres solutions, depending on whether $\beta,_r =
0$ or $\beta,_r \neq 0$. The first family is a simultaneous generalisation of
the Friedmann and Kantowski--Sachs models \cite{KaSa1966}. So far it has found
no application in astrophysical cosmology, and we shall not discuss it here. The
metric functions in the second family are
\begin{eqnarray}\label{e2}
{\rm e}^{\beta} &=& \Phi(t, r) {\rm e}^{\nu(r, x, y)}, \nonumber \\
{\rm e}^{\alpha} &=& h(r) \Phi(t, r) \beta,_r \equiv h(r) \left(\Phi,_r + \Phi
\nu,_r\right), \nonumber \\
{\rm e}^{- \nu} &=& A(r)\left(x^2 + y^2\right) + 2B_1(r) x + 2B_2 (r)y + C(r),
\end{eqnarray}
where $\Phi(t, r)$ is a solution of
\begin{equation}\label{e3}
{\Phi,_t}^2 = - k(r) + \frac {2 M(r)} {\Phi} + \frac 1 3 \Lambda \Phi^2;
\end{equation}
$\Lambda$ is the cosmological constant, while $h(r)$, $k(r)$, $M(r)$, $A(r)$,
$B_1(r)$, $B_2(r)$ and $C(r)$ are arbitrary functions obeying
\begin{equation}\label{e4}
g(r) \df 4 \left(AC - {B_1}^2 - {B_2}^2\right) = 1/h^2(r) + k(r),
\end{equation}
where $g(r)$ is another arbitrary function of the coordinate $r$ defined as
above. The mass density in energy units is
\begin{equation}\label{e5}
\kappa \rho = \frac {\left(2M{\rm e}^{3\nu}\right),_r} {{\rm e}^{2\beta}
\left({\rm e}^{\beta}\right),_r}; \qquad \kappa = 8 \pi G / c^4.
\end{equation}
The bang time function $t_B(r)$ follows from (\ref{e3}):
\begin{equation}\label{e6}
\int\limits_0^{\Phi}\frac{{\rm d} \widetilde{\Phi}}{\sqrt{- k + 2M /
\widetilde{\Phi} + \frac 1 3 \Lambda \widetilde{\Phi}^2}} = t - t_B(r).
\end{equation}

The Szekeres metric has in general no symmetry, but acquires a 3-dimensional
symmetry group with 2-dimensional orbits when $A$, $B_1$, $B_2$ and $C$ are all
constant.

The sign of $g(r)$ determines the geometry of the (constant $t$, constant $r$)
2-surfaces. The geometry is spherical, planar or hyperbolic (pseudo-spherical)
when $g > 0$, $g = 0$ or $g < 0$, respectively. With $A$, $B_1$, $B_2$ and $C$
being functions of $r$, the surfaces $r =$ const within a single space $t =$
const may have different geometries, i.e.\ they can be spheres in one part of
the space and surfaces of constant negative curvature elsewhere, the curvature
being zero at the boundary.

The sign of $k(r)$ determines the type of evolution when $\Lambda = 0$; with $k
> 0 = \Lambda$ the model expands away from an initial singularity and then
recollapses to a final singularity; with $k < 0 = \Lambda$ the model is either
ever-expanding or ever-collapsing; $k = 0 = \Lambda$ is the intermediate case
corresponding to the `flat' Friedmann model. Similarly to $g$, $k$ can have
different signs in different regions of the same space. The sign of $k(r)$
influences the sign of $g(r)$. Since $1/h^2$ in (\ref{e4}) must be non-negative,
we have the following: with $g > 0$ (spherical geometry), all three types of
evolution are allowed; with $g = 0$ (plane geometry), $k$ must be non-positive
(only parabolic or hyperbolic evolutions are allowed); and with $g < 0$
(hyperbolic geometry), $k$ must be strictly negative, so only the hyperbolic
evolution is allowed. The geometry of the latter two classes is poorly
understood \cite{HeKr2008,Kras2008}, and therefore not explored for cosmological
applications. Only the quasi-spherical model has been well investigated, and has
found applications in the study of the early Universe \cite{Gron1985,Moff2006},
structure formation \cite{Bole2006,Bole2007}, supernova \cite{BoCe2010} and
cosmic microwave background (CMB) \cite{Bole2009} observations, light
propagation \cite{KrBo2010}. In \cite{KrBo2010} it was shown that two rays sent
from the same source at different times to the same observer pass through
different sequences of intermediate matter particles. The change of object
position in the sky, due to this effect, should be observable in the future.

The quasi-spherical Szekeres models can be imagined as deformations of the
spherically symmetric models after which the spheres (still identifiable in the
Szekeres geometry) are no longer concentric. The mass-density distribution may
be interpreted as a superposition of a mass monopole and a mass dipole
\cite{DeSo1985, PlKr2006}.\footnote{A special case of this may be interpreted as
a pure mass dipole, but then the density is necessarily negative over approx.
half of each sphere, and the physical interpretation of such an object is
unknown.}

\medskip

{\underline {(2) The Lema\^itre and Lema\^itre -- Tolman models}}

\medskip

{\underline {(2.a) The Lema\^itre model}}

\medskip

The Lema\^itre metric \cite{Lema1933} describes a spherically symmetric
inhomogeneous fluid with anisotropic pressure.\footnote{The subclass of
isotropic pressure is usually credited to Misner and Sharp \cite{MiSh1964}, and
occasionally to Podurets \cite{Podu1964}.} In comoving coordinates it has the
following form
\begin{equation}
{\rm d}s^2 = {\rm e}^{A(t,r)} {\rm d}t^2 - {\rm e}^{B(t,r)}{\rm d}r^2 - R^2(t,r)
\left({\rm d}\vartheta^2 + \sin^2 \vartheta {\rm d}\varphi^2 \right). \label{e7}
\end{equation}
The Einstein equations reduce to:
\begin{equation}\label{e8}
 \kappa R^2 R,_r \rho = 2{M,_r},
\end{equation}
\begin{equation}
 \kappa R^2 {R,_t} p = -2 {M,_t},
\label{e9}
\end{equation}
where $\left(R,_t, R,_r\right) \df (\pdril R t, \pdril R r)$, $p$ is the
pressure, $\rho$ is the mass density in energy units, and $M(t,r)$ is defined
by:
\begin{equation} \label{e10}
2M = R + {\rm e}^{-A} R {{R,_t}}^2 - {\rm e}^{-B} R {R,_r}^2 -
\frac{1}{3} \Lambda R^3.
\end{equation}

In the Newtonian limit, $M c^2/G$ is equal to the mass inside the shell of
radial coordinate $r$. However, in curved space it is not an integrated rest
mass, but the active gravitational mass that generates the gravitational field.
As can be seen from (\ref{e9}), in the expanding Universe the mass decreases
with time. The function $B$ can be written in the following form \cite{Brad}
\begin{equation}
{\rm e}^{B(t,r)} = \frac{{R,_r}^2(t,r)}{1 + 2E(r)} \exp \left(
\int\limits_{t_0}^t {\rm d} \widetilde{t} \frac{2 {R,_t}(\widetilde{t},r)}{
\left[ \rho(\widetilde{t},r) + p(\widetilde{t},r) \right] R,_r(\widetilde{t},r)}
p,_r(\widetilde{t},r) \right), \label{e11}
\end{equation}
where $E(r)$ is an arbitrary function.
The equations of motion ${T^{\alpha \beta}}_{; \beta} = 0$ reduce to
\begin{eqnarray}
T^{0 \alpha}{}_{; \alpha} = 0 & ~~\Rightarrow & ~~B,_t + 4 \frac{{R,_t}}{R} = -
\frac{2 {\rho},_t}{ \rho + p},
\label{e12} \\
 T^{1 \alpha}{}_{; \alpha} = 0 & ~~\Rightarrow & ~~A,_r = - \frac{2 p,_r}{ \rho +
 p},
\label{e13} \\
T^{2 \alpha}{}_{; \alpha} = T^{3 \alpha}{}_{; \alpha} = 0 & ~~\Rightarrow &
~~\frac{\partial p}{\partial \theta} = 0 = \frac{\partial p}{\partial \phi}
\label{e14}.
\end{eqnarray}
The Lema\^itre model has been employed to study the conditions of the early
Universe \cite{GP89}, the mass of the Universe \cite{AH09}, supernova
observations \cite{LB10}, structure formation and the impact of pressure
gradients on shell crossing singularities \cite{BoLa2008}.

\medskip

{\underline {(2.b) The Lema\^itre-Tolman model}}

\medskip

In the special case of dust with the cosmological constant, the above equations reproduce the
Lema\^{\i}tre--Tolman (LT) model \cite{Lema1933,Tolm1934}.
When $p_{,r} = 0$,
equation (\ref{e13}) implies $A_{,r} = 0$, which means that the component
$g_{00}$ can be scaled to 1, and, using (\ref{e11}), the metric (\ref{e7})
becomes
\begin{equation}
{\rm d}s^2 = {\rm d}t^2 - \frac{R,_r^2}{1 + 2E} {\rm d}r^2 - R^2(t,r) \left({\rm
d}\vartheta^2 + \sin^2 \vartheta {\rm d}\varphi^2 \right).
\end{equation}
Equation (\ref{e10}) becomes then identical to (\ref{e3}):
\begin{equation}\label{e16}
 R_{,t}{}^2 = 2E + \frac{2M}{R} + \frac{\Lambda}{3} R^2.
 \end{equation}
Because the pressure is zero, the mass does not depend on time. The mass density
follows from (\ref{e8}), and the bang time function $t_B(r)$ is given by
(\ref{e6}), with $(-k, \Phi)$ replaced by $(2E, R)$.

For reviews of applications of the LT models see \cite{Kras1997, PlKr2006,
BKHC2009}. Selected examples: formation of black holes \cite{KrHe2004b,FR08}, of
galaxy clusters \cite{KrHe2002,KrHe2004a}, superclusters \cite{BolHel06}, cosmic
voids \cite{BoKH2005}, interpretation of supernova observations \cite{Cel00,
IgNN2002, AAG06, chung06, AlAm2007, enqvist07,BMN07, BTT07a, BTT07b, MKMR2007,
bolejko08, MK2008,BT08,GH08a,GH08b, CFL08,enqvist08,BN08,BoWy2009,EMR09}, CMB
\cite{ABNV2009, ZaMS2008,WV09,RC10,ClRe2010,MZS10}, redshift drift
\cite{UCE08,AS09}, and averaging \cite{BB1,RR1,WW1}.

Some of these applications will be discussed in Sec. \ref{DirMet}.

\medskip

{\underline {(3) The Stephani -- Barnes (S--B) family}}

\medskip

This is the family of perfect fluid solutions with zero shear, zero rotation and
nonzero expansion. It consists of two collections of solutions:

\medskip

{\underline {(3a) The conformally flat solution:}}

\medskip

\begin{equation}
ds^{2} = D^{2}dt^{2} - V^{-2}(t,x,y,z) (dx^{2} + dy^{2} + dz^{2}), \label{e17}
\end{equation}
where:
\begin{eqnarray}
\hspace{-2cm} D &=& F(t)V,_{t}/V, \label{e18} \\
\hspace{-2cm} V &=& \frac 1 R \left\{1 + \frac 1 4 k(t) \left[\left(x -
x_{0}(t)\right)^{2} + \left(y - y_{0}(t)\right)^{2} + \left(z -
z_{0}(t)\right)^{2}\right]\right\}, \label{e19}
\end{eqnarray}
$F(t), R(t), k(t), x_{0}(t), y_{0}(t)$ and $z_{0}(t)$ are arbitrary functions of
time, $F$ is related to the expansion scalar $\theta $ by $\theta = 3/F$, and
$k(t)$ is a generalisation of the FLRW curvature index $k$, it can change sign
during evolution. The matter density and pressure are:
\begin{equation}
\kappa \rho c^2 = 3kR^{2} + 3/F^{2} \df 3C^{2}(t), \label{e20}
\end{equation}
\begin{equation}
\kappa p = - 3C^{2}(t) + 2CC,_{t}V/V,_{t} . \label{e21}
\end{equation}
This solution was found by Stephani \cite{Step1967}; it is the most general
conformally flat solution with a perfect fluid source and nonzero expansion. As
seen from (\ref{e20}) --  (\ref{e21}), the matter density in it depends only on
the comoving time, while the pressure depends on all the coordinates. In
general, the solution has no symmetry. In Refs. \cite{BoCo1985, BoCo1988,
CoFe1989, KQSu1997} it was shown that the source has the thermodynamics of a
single-component perfect fluid only if the metric (\ref{e20}) -- (\ref{e21}) is
specialised so that it acquires an at least 3-dimensional symmetry group acting
on at least 2-dimensional orbits. The FLRW limit follows when the functions $k,
x_{0}, y_{0}$ and $z_{0}$ are all constant.

The arbitrary functions of time cause that the evolution of the spacetime is not
determined. This is because no equation of state was imposed on (\ref{e20}) --
(\ref{e21}). Unfortunately, the two types of equations of state that are most
often used in cosmology and astrophysics (dust, $p = 0,$ and a barotropic
equation of state, $f(p,\rho ) = 0)$ both reduce  (\ref{e20}) -- (\ref{e21}) to
an FLRW model.

\medskip

{\underline {(3b) The Petrov type D solutions}}

\medskip

Equations (\ref{e20}) and (\ref{e21}) still apply here, but now $V(t, x, y, z)$
is determined by the following equation (resulting from the Einstein equations):
\begin{equation}
w_{uu} /w^{2} = f(u), \label{e22}
\end{equation}
where $f(u)$ is an arbitrary function. The variable $u$ and the function $w$ are
related to the coordinates $x, y, z$, and to the function $V(t,x,y,z)$ as
follows:
\begin{equation}\label{e23}
\hspace{-2cm} (u, w) = \left\{ \begin{array}{lll} (r^{2}, V)
   & \mbox{for spherically symmetric models};\\
(z, V)  & \mbox{for plane symmetric models};\\
(x/y, V/y)  & \mbox{for hyperbolically symmetric models},\end{array} \right.
\end{equation}
where $r^2 = x^2 + y^2 + z^2$. These three classes of models were found by
Barnes \cite{Barn1973}, but the spherically symmetric case was known much
earlier (and rediscovered many times over, see \cite{Kras1997} for a full list).
The Einstein equations were reduced to the form (\ref{e22}) by Kustaanheimo and
Qvist \cite{KuQv1948}. With $f(u) = 0,$ the Barnes models all become conformally
flat and are then subcases of the Stephani solution.

Many papers discussed methods of solving (\ref{e22}) and examples of particular
solutions, but with no relation to cosmology. An interesting application was a
counterexample to the Ehlers -- Geren -- Sachs (EGS) theorem
\cite{ClBa1999,BaCl2000} -- it was shown that almost isotropic CMB is also
possible in an inhomogeneous Universe.

One member of the Barnes family of solutions, found by McVittie \cite{McVi1933},
is worth noting here:
\begin{eqnarray}\label{e24}
{\hspace{-2cm}} && {\rm d}s^2 = \left[\frac {1 - \mu (t,r)} {1 + \mu
(t,r)}\right]^2 {\rm d}t^2 - R^{2}(t) \frac {[1 + \mu (t,r)]^4} {\left(1 + \frac
1 4\ kr^2\right)^2} \left[{\rm d}r^2 + r^2 \left({\rm d}\vartheta^2 + \sin^2
\vartheta {\rm d} \varphi^2\right)\right], \\
{\hspace{-2cm}} && {\rm where:} \qquad \mu (t,r) = \frac m {2rR} \sqrt{1 + \frac
1 4\ k r^2}, \nonumber
\end{eqnarray}
$m$ and $k$ being arbitrary constants and $R(t)$ being an arbitrary function. In
the case $m = 0$ this solution reproduces the whole FLRW class, and when $(k, R)
= (0, 1)$ it reproduces the Schwarzschild solution. So, it is an exact
superposition of the FLRW and Schwarzschild metrics, with a perfect fluid
source. It was published in 1933 (!).

A few authors attempted to apply this solution to observational cosmology.
However, all those attempts were fallacious. McVittie's discussion of the
influence of cosmic expansion on planetary orbits was coordinate-dependent.
J\"{a}rnefelt's perturbative discussions of the same problem \cite{Kras1997}
produced no conclusive result because the author did not define a length unit
that would be unchanging in time. Later, McVittie \cite{McVi1966} applied his
solution to a discussion of stellar collapse, but the case $k = 0$ that he
discussed has a spatially homogeneous density, so is unrealistic. Noerdlinger
and Petrosian \cite{NoPe1971} considered the problem of whether clusters or
superclusters of galaxies participate in the cosmological expansion. Their
discussion was mostly Newtonian; they used the McVittie solution only to
estimate the relativistic correction to the result.

The disadvantage of this solution, shared with the whole S--B family, is that it
contains an arbitrary function of time, and so does not define any evolution law
for the Universe. One way out of this is to impose an equation of state -- but
so far no-one had a workable idea on what this equation should be. A barotropic
equation $f(\rho,p) = 0$ reduces the McVittie solution to pure FLRW.

The subcase $k = 0$ is of little interest for cosmology because, similarly to
the Stephani \cite{Step1967} solution, it has spatially homogeneous mass
density, and so its whole inhomogeneity is hidden in pressure gradients. So far,
nobody has provided a physical interpretation of this situation.

Global properties of the McVittie solution were discussed by Sussman
\cite{Suss1988}. He showed that the seemingly self-evident interpretation (a
point particle in an expanding universe) is not consistent with the global
geometry. The set $r = 0$ is a null boundary, and its intersection with any $t
=$ constant hypersurface $H$ is at an infinite geodesic distance from any other
point of $H$.

More recently, there appeared a collection of papers discussing global
properties of the McVittie solution, and some generalisations of it, for example
by Nolan \cite{Nola1998, Nola1999}, Carrera and Giulini \cite{CaGi2010} and
Kaloper {\it et al.} \cite{KKMa2010}. However, from the point of view of
cosmology, the results of these papers would require clarification, for which
there is not enough space in this review, for the following reasons:

1. They discuss the physically uninteresting subcase $k = 0$.

2. They treat the McVittie solution as if it were the only existing candidate
for a model of a black hole embedded in an FLRW universe. They overlook the fact
that it is a member of the large Barnes family that might be surveyed for more
such examples. They also overlook the fact that the LT and Szekeres models do
contain subcases describing black holes in a cosmological background, which are
physically much better understood, see for example Ref. \cite{KrHe2004b}.

3. They are involved in a tangle of polemics, the later authors pointing out
alleged errors in the earlier papers. Consequently, an extended re-analysis
would be necessary in order to sort out who is right.

4. Some errors in the most recent papers are evident. Examples:

(4a) Carrera and Giulini \cite{CaGi2010} cite Sussman \cite{Suss1988} and
Gautreau \cite{Gaut1984} as examples of a ``confusion'' about interpreting the
McVittie solution as a point particle in an FLRW universe. In truth, Sussman was
the first to point out {\em and resolve} this confusion, while Gautreau's paper
has nothing to do with McVittie (it used the LT model to discuss the influence
of cosmic expansion on planetary orbits).

(4b) Kaloper {\it et al.} \cite{KKMa2010} claim that there is some kind of
singularity where invariants built of {\em second derivatives} of the Riemann
tensor diverge; they call it a ``very soft'' singularity. In this, they revive
the infamous ``weak singularity'' concept of Vanderveld {\it et al.}
\cite{VFWa2006}, which was proven in Refs. \cite{BKHC2009} and \cite{KHCB2010}
to be no singularity at all; see also sec. 7 of the present text.

\medskip

{\underline {(4) Generalisations of the LT and Barnes models}}

\medskip

For the LT model and for the whole Barnes class generalisations were found in
which the matter source is a charged dust, or, respectively, a charged perfect
fluid obeying the Einstein -- Maxwell equations. These do not seem to have a
direct application in cosmology, so we do not review them here; see Refs.
\cite{Kras1997} and \cite{PlKr2006} for overviews. Nevertheless, the charged LT
solution has interesting physical properties \cite{PlKr2006, KrBo2006,
KrBo2007}. In addition, several generalisations of the LT and Barnes models were
found, in which the source has nonzero viscosity or heat conduction. The
physical interpretation of these in a cosmological context is less clear; see
Ref. \cite{Kras1997} for a review.

\medskip

{\underline {(5) Other models}}

\medskip

The list that might be given here depends on how one defines a cosmological
model. In Ref. \cite{Kras1997} it was proposed that the term ``cosmological
model'' may denote only such a solution of Einstein's equations that contains a
nontrivial member of the FLRW class as a limiting case. We shall stick to this
terminology here, thereby eliminating more than 1500 papers \cite{Kras1997}
whose authors used the term ``cosmological model'' for their results. In Ref.
\cite{Kras1997} this definition was used in a strict formal way, which resulted
in listing a large number of metrics, most of which do not seem to have any
relation to observational cosmology because, for example, they contain fields of
unclear interpretation, often coupled together in ways that are difficult to
interpret. Examples of those are briefly listed here to give the reader an idea
about the wealth of the existing material.

\medskip

{\underline {(5a) Models with null radiation}}

\medskip

These are superpositions of the FLRW models with various vacuum solutions, like
those of Schwarzschild, Kerr, Kerr -- Newman, etc. The superpositions are not
perfect fluid solutions, and their energy-momentum tensors were interpreted ex
post as mixtures of perfect fluid with null radiation (whose energy-momentum
tensor is $T_{\mu \nu } = \tau k_{\mu }k_{\nu }$ with $k^{\mu }k_{\mu } = 0)$,
sometimes also with electromagnetic field. The solutions were in fact guessed in
the course of exercises in metric-building and interpreting. As a result, the
different contributions to the source are coupled through common constants so
that, for example, the null radiation can in some cases vanish only if either
the perfect fluid component or the inhomogeneity on the FLRW background go away.
In particular, the superposition of the Schwarzschild and FLRW solutions in this
family is different from the McVittie solution \cite{McVi1933}. This activity
was started by Vaidya \cite{Vaid1977}, who found a superposition of the Kerr and
FLRW solutions, and the probably most sophisticated composite was found by Patel
and Koppar \cite{PaKo1988}; it is an infinite sequence of perturbations of the
flat FLRW background whose first-order term is the Kerr solution.

\medskip

{\underline {(5b) The ``stiff-fluid'' models}}

\medskip

These are solutions of the Einstein equations with a 2-dimensional Abelian
symmetry group acting on spacelike orbits, in which the perfect fluid source
obeys the ``stiff equation of state'', energy density = pressure (the source can
be alternatively interpreted as a massless scalar field). It was claimed that
these models apply to the early Universe, but the real reason behind the
popularity of this activity was that such solutions can be relatively simply
generated from vacuum solutions with the same symmetry, of which many are known.
This activity began with the paper by Tabensky and Taub \cite{TaTa1973}, and the
probably most sophisticated example of an explicit solution was given by
Belinskii \cite{Beli1979}. See \cite{Kras1997} for an extended review.

\medskip

{\underline {(5c) Examples of other solutions (see \cite{Kras1997} for a full
list)}}

\medskip

1. The Petrov type $N$ perfect fluid solutions of Oleson \cite{Oles1971}.

2. A few simple examples of spherically symmetric perfect fluid solutions with
shear, expansion and acceleration being all nonzero, see Ref. \cite{Kras1997}.

3. Examples of algebraically special solutions defined by requirements imposed
on the degenerate principal null congruence of the Weyl tensor \cite{Kras1997}.

4. Anisotropic soliton-like perturbations propagating on the flat FLRW
background (the pressure has different values for different directions). The
most elaborate example of an explicit solution was given by Diaz, Gleiser and
Pullin \cite{DiGP1988}.

In this article we will discuss only those exact inhomogeneous cosmological
models that allow for testable observational predictions. For more exhaustive
discussions the reader is referred to \cite{Kras1997,PlKr2006,BKHC2009}.

\section{Distance measurements}\label{dist}

The concept of distance lies at the root of almost all cosmological observations
whose interpretation strongly depends on this quantity. The distance however
depends on the assumed model of the Universe and on the matter distribution in
it. The effect of inhomogeneities on the measured distance has been addressed
frequently after the papers by Kristian \& Sachs \cite{KrSa1966} and Dyer \&
Roeder \cite{DyRo1973} were published (see also \cite{Rasa2009} and references
therein). For example, Partovi \& Mashhoon \cite{PaMa1984} showed that the
inhomogeneities affect the second order coefficient in the series expansion of
the luminosity distance, i.e. the deceleration parameter. Using the same line of
calculations Pascual-S\'anchez argued that in such a case the deceleration
parameter can be negative just due to the presence of inhomogeneities
\cite{PaSa1999}. However, cosmologists often disregard the effect of
inhomogeneities and just apply the FLRW relation. The `justification' is: 1)
even if density variations are large, the fluctuations of the gravitational
potential are small and therefore the perturbation scheme can be applied, and 2)
since the perturbations are Gaussian, they vanish after averaging, and therefore
they should have little impact on observations. However, as shown by Sachs
\cite{Sach1961}, the equation for the angular diameter distance $D_A$ is
\begin{equation}
\frac{{\rm d^2} D_A}{{\rm d} s^2} = - ( |\sigma|^2 + \frac{1}{2} R_{\alpha
\beta} k^{\alpha} k^{\beta}) D_A, \label{e25}
\end{equation}
where $\sigma$ is the shear, $R_{\alpha \beta}$ is the Ricci tensor and
$R_{\alpha \beta} k^\alpha k^\beta = \kappa T_{\alpha \beta} k^\alpha k^\beta$.
In the case of dust ($p=0$), in the comoving and synchronous coordinates,
$R_{\alpha \beta} k^\alpha k^\beta = \kappa \rho k^0 k^0$.
As seen, the distance {\em does} depend on density fluctuations (not on the
gravitational potential), and secondly, even if the perturbations vanish after
averaging (i.e. $\av{\rho} = \av{\rho_b + \delta \rho} = \rho_b$, where $\rho_b$
is the background density), they do modify the distance and the final result
deviates from the homogeneous solution $\rho = \rho(t)$. This is a consequence
of (\ref{e25}). This means that one needs to know an exact model to calculate
the distance -- a statistical information about the density distribution, like
the matter power spectrum, is not sufficient to calculate it. The matter power
spectrum can only be used (within the linear regime) to estimate fluctuations
around the mean distance-redshift relation, to be precise $\av{\Delta_D^2}$
(where $\Delta_D$ is given by (\ref{deldis})). Thus, this method does not
provide any information about the change of $\Delta_D$, which as mentioned
above, very often is assumed to be zero. Apart from the above mentioned 2
arguments sometimes people quote Weinberg's argument \cite{Wein1976} that
although for a single case the distance is modified by the inhomogeneities, but
due to photon conservation, when averaged over large enough angular scales the
overall effect is zero. A detailed discussion why this kind of reasoning should
not apply is presented in \cite{ElBD1998}.

\begin{figure}
\begin{center}
\includegraphics[scale=0.8]{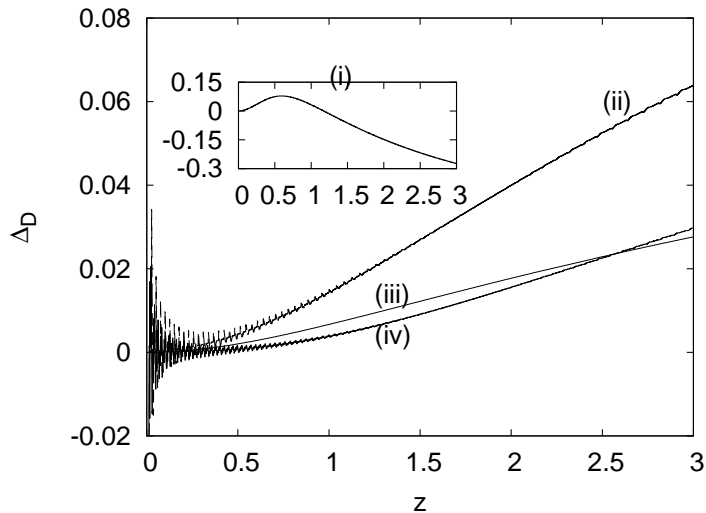}
\caption{The distance correction. Roman numbers refer to the labels of the
models discussed in this section. Models (ii)-(iv) are based on the Swiss-Cheese
models with Mpc-scale inhomogeneities hence large fluctuations for low
redshifts.} \label{discor}
\end{center}
\end{figure}

To show how matter inhomogeneities affect the distance let us consider the
following examples:
\begin{enumerate}
\item A large-scale inhomogeneous matter distribution (Gpc-scale) whose volume
average does not vanish,
$\av{\delta \rho} \ne 0$. This
is the giant void model with best fit parameters as presented in \cite{BoWy2009}
(the reader is referred there for more details).
\item Small-scale inhomogeneities (Mpc-scale)
whose volume average
vanishes,
$\av{\delta \rho} = 0$. However, the average of the density fluctuations along
the line of sight is not zero $\av{\delta \rho}_{1D} \ne 0$. The model is based
on the Swiss-Cheese model presented in \cite{Bole2011} and the reader is
referred there for more details.
\item
As above but now
the distance is calculated within the weak lensing approximation, and is based
on the model presented and discussed in \cite{Bole2011a}.

\item Small-scale inhomogeneities (Mpc-scale)
whose volume average
vanishes,
$\av{\delta \rho} = 0$. Also, the average of density fluctuations along the line
of sight is zero, $\av{\delta \rho}_{1D} = 0$. The model is based on the
Swiss-Cheese model presented in \cite{Bole2011b}.
\end{enumerate}
The results in terms of the distance corrections, $\delta_D$, are presented in
Fig. \ref{discor}. The distance correction $\delta_D$ is defined by the
following relation
\begin{equation}
D_A = \bar{D}_A (1 + \Delta_D),
\label{deldis}
\end{equation}
where $\bar{D}_A$ is the distance in the homogeneous (background) model. As
seen, the correction is of the order of a few percent, thus, owing to the
increasing precision of the observations, the inhomogeneities need to be taken
into account.

Below we will discuss several examples showing how inhomogeneities can influence
our interpretation of cosmological observations. As we cannot discuss here every
single paper that deals with this issue we will just focus on some major
developments and refer only to a few (not all) papers dealing with this problem.
We will also omit the Stephani models in our review as being less physically
motivated.

\section{The direct method}\label{DirMet}

The studies of inhomogeneities and their effect on observations can be divided
in 3 approaches: direct methods, the inverse problem and the averaging approach.
Only the first two will be discussed in this review -- for the averaging
approach see \cite{BB1,RR1,WW1}. In the first approach a model is specified by a set of a priori chosen
parameters and the observational data is used to find the best fit for these
parameters. In the second approach the observational data is used to define the
model with no a priori constraints imposed on it.

\subsection{Giant void models}\label{giantvoid}

The giant void configurations are characterised by underdensity profiles
increasing with radius on Gpc scales. One of the first and simplest models was
the one discussed by Tomita \cite{Tomi2000,Tomi2001a,Tomi2001b}. He considered a
model consisting of a low-density inner and a higher density outer homogeneous
regions connected at some redshift and showed that such a configuration can
explain the supernova dimming. After 2006 the number of papers concerning the
giant void models rapidly increased.
Assuming a density profile and an expansion rate, or a shape of the bang time
function, one analyses cosmological observations to constrain the parameter
space of the giant void. However, the particular constraints strongly depend on
the assumed parameterisations, and almost every single paper introduced its own.
As it is impossible to discuss all of them, we will focus here on three
examples:

\begin{enumerate}
\item The GBH void model \cite{GH08a} is defined by the following functions
\begin{equation}\label{e27}
M(r) =  \frac{1}{2} H_0(r)^2 \Omega_m(r) R_0^3 \quad \& \quad k(r) = H_0(r)^2 (
\Omega_m(r) - 1) R_0^2,
\end{equation}
where
\begin{eqnarray}
&&\Omega_m(r) = \Omega_{\rm out} + \Big(\Omega_{\rm in} - \Omega_{\rm out}\Big)
\left({1 - \tanh[(r - r_0)/2\Delta r]\over1 + \tanh[r_0/2\Delta r]}\right),\\
&&H_0(r) =  H_{\rm out} + \Big(H_{\rm in} - H_{\rm out}\Big) \left({1 - \tanh[(r
- r_0)/2\Delta r]\over1 + \tanh[r_0/2\Delta r]}\right).
\end{eqnarray}
There are 6 parameters here, $\Omega_{\rm out}$ determined by the assumption of
asymptotic flatness, $\Omega_{\rm in}$ determined by LSS observations, $H_{\rm
out}$ determined by CMB observations, $H_{\rm in}$ determined by HST
observations, $r_0$ characterizing the size of the void, $\Delta r$
characterizing the transition to uniformity. But in the GBH model it is assumed
that $\Omega_{\rm out} = 1$.

\item Bolejko and Wyithe class I model \cite{BoWy2009} is defined by

\begin{equation}\label{e30}
\rho(t_0,r) = \rho_b \left[ 1 + \delta_{\rho} - \delta_{\rho} \exp \left( -
\frac{r^2}{\sigma^2} \right) \right], \quad \& \quad t_B = 0.
\end{equation}
where $\rho_b = \Omega_m  \times (3H_0^2)/(8 \pi G)$. It contains 4 parameters:
$H_0, \Omega_m, \delta_{\rho}, \sigma$. In \cite{BoWy2009} it was assumed that
$\Omega_m=0.3$ but here we will allow this parameter to vary.

\item The spline model of Zibin, Moss, and Scott \cite{ZaMS2008,MZS10}
is defined by
\begin{equation}\label{e31}
\rho(t_0,r) = \rho_{EdS} ( 1 + \delta), \quad \& \quad t_B = 0,
\end{equation}
where $\delta$ is given by a three-point cubic spline to the initial density
fluctuation $\delta_{j} = \delta_{t_i,r_j}$ = and $j = 1, 2, 3$. By
construction, $r_1 = 0 = \delta_3$. Thus the model depends on 5 free parameters:
$\delta_1, \delta_2, r_2, r_3$ and $\rho_{EdS}$ which just depends on $H_0$.
\end{enumerate}

\begin{figure}
\begin{center}
\includegraphics[scale=0.8]{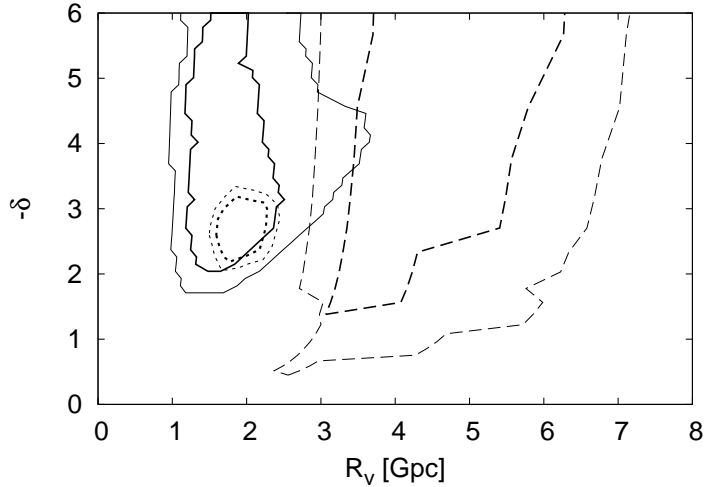}
\caption{Supernova constraints on the size and depth of giant void models.
Dashed line: the GBH model (\ref{e27}), solid line: the BW model (\ref{e30}),
and dotted line: the ZMS spline model (\ref{e31}).} \label{SNcon}
\end{center}
\end{figure}

Using cosmological observations (like supernovae and CMB, etc.) one can
constrain the range of the above parameters. Usually three parameters are most
interesting: the size and depth of the void and the local value of $H_0$. In
order to compare the constraints from these 3 parameterisations let us assume
that the depth is just the density contrast ($\delta = 1 - \rho_{out}/
\rho_{in}$) and the radius of a void is defined as a place (at the current
instant) where the density contrast becomes smaller than $-0.1$ as we proceed
from outside into the void.

The constraints on $\delta$ and $R_v$ from the supernova observations\footnote{
The constraining function is $\chi^2 = \sum_i \frac{(\mu_i -
\mu_0)^2}{\sigma_i^2}$, where $\mu_i$ and $\sigma_i$ correspond to the
measurements of the 557 supernovae \cite{ALRA2010}, $\mu_0$ is the distance
modulus in the considered model.} are presented in Figure \ref{SNcon}. As seen,
using different parameterisations one obtains different constraints. For
example, a void of size $R_v = 1.5$ Gpc and $\delta = -3$ is consistent with the
constraints coming from (\ref{e30}), but is excluded by those coming from
(\ref{e27}).

Apart from supernovae it is common to include the CMB constraints. However, up
to date no one has performed the full CMB analyses within the LT framework. In
the standard approach (the FLRW framework) the CMB data is analysed using the
temperature anisotropy power spectrum given by the covariance of the temperature
fluctuation expanded in spherical harmonics

\begin{equation}
        C_l = 4\pi \int \frac{dk}{k} {\cal P}_i |\Delta_l(k,\eta_0,\mu)|^2\,
\end{equation}
where $\Delta_l(k,\eta_0,\mu)$ is the transfer function, ${\cal P}_i$ is the
initial power spectrum, $\eta_0$ is the conformal time today and $\mu$ is the
angle $\mu = {\bf k.n}/k$ (with $\bf{n}$ the unit vector in the direction of the
emission of the radiation). On large scales the transfer function is of the form
$ \Delta_l(k,\eta_0) = \Delta_l^{\rm LSS}(k) +  \Delta_l^{\rm ISW}(k)$, where
$\Delta_l^{\rm LSS}(k)$ is the contribution from the last scattering surface
given by the Sachs-Wolfe effect and the temperature anisotropy, and
$\Delta_l^{\rm ISW}(k)$ is the contribution due to the change in the
gravitational potential along the line of sight, known as the integrated
Sachs-Wolfe (ISW) effect. The ISW effect depends on the growth of the
perturbations within the considered model. As the perturbative scheme within the
LT model is still in its infancy \cite{Zibi2008,ClCF2009,CC1} it is impossible to estimate the ISW
effect in the conventional way. However, the ISW effect is only important for
low $\ell$ and is expected to be smaller than the cosmic variance $\Delta C_l /
C_l = \pm (2/(2 \ell + 1))^{1/2}$. Another effect that is hard to estimate, but
expected to be within the cosmic variance limits, is the effect of the
reionisation.

Therefore, the analysis of the CMB within the LT framework is done as follows:
it is assumed that the generation of the CMB anisotropies at the last scattering
is the same as in the standard model. As the post-decoupling effects (like the
ISW effect or reionisation) are expected to be smaller than cosmic variance, one
does not estimate these effects within the LT framework, but uses standard codes
like CMBFAST \cite{SZ96} or CAMB \cite{CAMB}. Therefore, if one does not change
the initial power spectrum then the shape and amplitude of the Doppler peaks is
just governed by $\Omega_c h^2, \Omega_b h^2, \Omega_k h^2$ (i.e., the physical
densities of cold dark mater, baryonic matter and curvature)
\cite{BoET1997,EfBo1999}. Finally, as the angular diameter distance maps the
physical position of the peaks to peaks in the angular power spectrum $C_l$ as a
function of the multipole $\ell$, one needs to fit the angular diameter distance
to the last scattering surface. Thus, the only difference between the standard
analysis of the CMB and the analysis within the LT model is the change of
distance to last scattering. Other effects have not been taken into account
because of lack of a fully developed perturbative scheme within the LT
framework, though the change of initial power spectrum was considered in
\cite{MZS10,NaSa2010}.

{}From the above description it is apparent that such an analysis only weakly
constrains the giant void \cite{ClRe2010}. To successfully fit the CMB data one
just needs to fit $\Omega_c h^2, \Omega_b h^2, \Omega_k h^2$ (in the region that
emitted CMB) and the distance to the last scattering surface. This however, as
argued in \cite{BoWy2009}, can be achieved by changing the properties of the
model outside the void. The region from which the CMB was emitted is currently
approximately 13 Gpc away from us, therefore its $\Omega_c h^2, \Omega_b h^2,
\Omega_k h^2$ are not related to a void which has a radius of $\sim 3$ Gpc.
Similarly, the distance to the last scattering instant can be tuned by the
properties of the model outside the void.

In most cases, however, one does not consider any modification outside the void
and just assumes that the Universe (from our Galaxy up to the last scattering)
is described by the chosen parametrisation, such as for example (\ref{e27}) or
(\ref{e31}). Such a procedure leads to large systematics. For example, if the
background (i.e the model far away from the void) is assumed to be the
Einstein-de Sitter model then in order to have a good shape of the CMB power
spectrum a low value of the expansion rate is required. This is because the
proper shape of the power spectrum requires that  $\Omega_m h^2 \approx 0.13$,
so if one assumes  that $\Omega_m =1$ then one gets $h \approx 0.4$. This, on
the other hand, has strong implications for the void. To fit the supernovae one
needs a fluctuation of the expansion rate of amplitude $\delta_H \approx 0.1 -
0.2$ \cite{enqvist07,BoWy2009}, so this implies that the local expansion rate is
low, i.e. $H_0 \approx 45$ km s$^{-1}$ Mpc$^{-1}$ \cite{MZS10} or $H_0 \approx
60$ km s$^{-1}$ Mpc$^{-1}$ \cite{GH08a}. This, when combined with local
measurements of $H_0$, seems to rule out the giant void. The assumption of an
Einstein-de Sitter background also impairs the BAO analysis like the one in
\cite{MZS10}. When the assumption of spatial flatness is relaxed one obtains
better results, see for example \cite{BNV10}.

The local  expansion rate within the LT region is also important for the age
considerations. A small $H_0$ implies a large age of the Universe \cite{MZS10}.
On the contrary, a large $H_0$ and $\Lambda = 0$ imply a shorter age. Thus $H_0
\approx 70$ km s$^{-1}$ Mpc$^{-1}$ gives the age of $11-12$ Gyr. This was
discussed in \cite{LaLW2010}, where it was shown that the current measurements
do not put tight constraints on the model.

Actually, in an LT model, due to the shear, the anisotropy in the expansion
increases as the void grows and becomes nonlinear. Hence, physical length
scales, and in particular the sound horizon at the drag epoch, which are
isotropic in a FLRW model, become more and more different in the radial and
transverse direction. This is the reason why the Einstein-de Sitter background
is suspected not to be a good approximation for the calculation of the BAO in
the case of huge (Gpc) voids.

Different tests have been proposed in the literature to rule out or confirm the
giant void proposal. The stronger constraints compared to other observational
probes are the kinematic Sunyaev-Zeldovich (kSZ) effects \cite{SuZe1980} that
can be observed on distant galaxy clusters. Since these clusters are off the
centre of the LT universe, there should be a large CMB dipole in their frame of
reference which would manifest itself for us as a kSZ effect. Such a test was
performed by the authors of \cite{GH08b} who showed that observations of nine
clusters with large error bars can rule out LT models with voids of size greater
than $\sim 1.5$ Gpc. More recently, using the observational data from the the
South Pole Telescope and the Atacama Cosmology Telescope, Zhang \& Stebbins
\cite{ZhSt2010} put tighter constrains on the size of a void $\sim 850$ Mpc.
However, the paper uses the ``Hubble bubble'' void model (i.e. based on a
2-region FLRW model, with negative curvature inside and spatially flat outside).
Also, if one introduces non-adiabatic perturbations then the observational
constraints are  relaxed \cite{YNS10}.

Another test uses the spectral distortion of the CMB black-body spectrum
\cite{CS08}. A large local void causes ionised gas to move outward, in motion
relative to the frame of the CMB. This produces a Doppler anisotropy in the
frame of the gas. A large void will imply large anisotropies which will be
reflected back at us as spectral distortions. The test has been performed in
\cite{CS08}, using the particular case of ``Hubble bubble'' void models. Large
voids with large density contrasts are thus ruled out. However, the test has
only been applied to the Hubble bubble class of models and other models may
evade this test.

The void can also be tested directly by means of galaxy surveys as the one
reported in \cite{KTBC2010}. Here, a deep, wide-field near-infrared survey is
presented and explored to provide implications for local large scale structure.
The results suggest that local structures may exist on scales up to 300 Mpc.

In the papers cited above, the observer has been assumed to be located at the
centre of the LT model. But one can find in the literature some models where
he/she is assumed to be off the centre. However, the CMB low multipoles put
stringent limits on the distance he/she can be from the centre. This has been
studied in \cite{AlAm2007,BlMo2010,JS99} using different LT models. Using SN Ia
data alone, it can be concluded that the observer can be displaced at most 15\%
of the void scale radius from the centre \cite{AlAm2007,BlMo2010}. But when one
takes into account the induced anisotropies in the CMB temperature, the
combination of the CMB dipole measurement and the SNe Ia data imposes very
strict constraints on how far from the centre the observer can be located, i. e.
no more than 1\% of the void scale radius \cite{BlMo2010}.

For more details on observational constraints on giant void models see 
\cite{ZZ1,MM1}.

\subsection{Non-void models -- the effect of expansion}

Giant voids are not the only configurations that can be used to fit cosmological
observations. There is a group of LT models that are defined by the assumption
of a homogeneous density distribution at the present instant and a Gpc-scale
inhomogeneous expansion rate. Such a configuration was first considered in
\cite{bolejko08}. A homogeneous density profile at the current instant does not
imply a homogeneous profile at all times. Also the bang time function  in such
cases is of high negative amplitude, around 1-2 Gyr at 2 Gpc \cite{bolejko08}.
The influence of inhomogeneous expansion on the luminosity distance was further
studied in more detail by Enqvist and Mattsson \cite{enqvist07}. In their set of
different LT models the observer is located at the centre and the universes are
defined by an inhomogeneous expansion rate and a homogeneous density profile in
some cases and in other cases by an inhomogeneous expansion and homogeneous
$t_B$.

The analysis of cosmological observations within this type of models implies
that they have a better goodness of fit than giant void models
\cite{enqvist07,BoWy2009}. The amplitude of the fluctuations of the expansion
needed to fit the observations was found to be around $\delta_H \approx 0.1 -
0.2$ \cite{enqvist07,BoWy2009}. This seems to imply that an inhomogeneous
expansion rate is very important, since, as shown in  \cite{bolejko08}, models
with a homogeneous expansion rate and $\Lambda =0$ cannot successfully fit
supernova data.

This property can be linked to the result of Krasi\'nski and Hellaby
\cite{KrHe2004a,BoKH2005} that velocity perturbations are more efficient at
generating structures than density perturbations.

\subsection{Swiss-Cheese models}

An alternative way of modelling inhomogeneities is the Swiss-Cheese approach.
Instead of assuming that the whole Universe is modelled by a single
inhomogeneity of Gpc-scale, smaller inhomogeneous patches that are matched with
each other are considered.

The Einstein and Straus \cite{EnSt1945} type Swiss-cheese models were used to
study, among other effects, the influence of inhomogeneities on the
magnitude-redshift relation \cite{Kant1969,Nott1982a,Nott1982b,Nott1983}.
However, since Schwarzschild's is a static solution, any influence of the
expansion of the vacuoles remains weak in such models, and the magnitude of the
reported effects is very low (as an example, Nottale \cite{Nott1982b}, using a
very simplified such model, found an observable amplification by medium density
clusters or by superclusters of galaxies of only a tenth of a magnitude).

The recent appearance in the literature of models of the Universe in which the
inhomogeneities are represented by LT regions within a homogeneous background,
where the matter is assumed continuously distributed, with densities both below
and above the average, allows us to account for this vacuole expansion. The
first authors, to our knowledge, to have considered such LT Swiss-cheese models
to deal with the dark energy problem were Kai et al. \cite{KaKo2007}. However,
their aim was to reproduce an accelerated expansion (which is only an artifact
of the homogeneity assumption), and not the observed luminosity
distance-redshift relation. Therefore, the constraints they found on their model
cannot be considered as relevant for cosmological purpose.

Other LT Swiss-cheese models have been proposed to deal with the same dark
energy problem \cite{BTT07a,BTT07b,BN08,MKMR2007,MK2008}. The best results were
obtained by Marra et al. \cite{MKMR2007,MK2008}, who considered a model where
holes with radius 350 Mpc are inserted into an Einstein-de Sitter background.
Each hole exhibits a low density interior, surrounded by a Gaussian density peak
near the boundary that matches smoothly to the exterior Friedmann density, and
such that the matter density in the centre is roughly $10^4$ times smaller than
in the Friedmann background. To have a realistic evolution, it is also demanded
that there are no initial peculiar velocities. This implies $0 < E(r) \ll 1$.
Evolving this model from the past to the present day, the inner almost empty
region expands faster than the background, and the interpolating overdense
region is squeezed by it. The density ratio between the background and the
interior of the hole increases by a factor of 2. The evolution is realistic.
Matter is falling toward the peaks in density. Overdense regions start
contracting and become thin shells, mimicking structures, while underdense
regions become larger, mimicking voids, and eventually they occupy most of the
volume. The propagation of photons is studied in three cases: the observer is
just outside the last hole, in the Friedmann region, looking at photons passing
through the holes; the observer is on a high density shell; the observer is in
the centre of a hole. The observables calculated are the redshift $z(\lambda)$,
the angular diameter distance $D_A(z)$, the luminosity distance $D_L(z)$ and the
corresponding distance modulus $\Delta m(z)$. In this model, inhomogeneities are
able to mimic at least partly the effects attributed to dark energy.

The last scenario  described above has some similarity to the one considered
years ago by Sato and coworkers \cite{MSS83,SM83,MS83a,MS83b,HS84}. Maeda,
Sasaki and Sato \cite{MSS83} considered a spherical void represented by a
low-density FLRW region surrounding the centre of symmetry, itself surrounded by
a LT transition region, in turn surrounded by a FLRW background with a higher
density which has positive curvature and recollapses. The void has a tendency to
expand forever, but it is eventually swallowed up in the final singularity of
the background FLRW region. Sato and Maeda \cite{SM83} have shown that spherical
symmetry is a stable property in the expansion of voids, i.e. initially
nonspherical voids become more spherical during their expansion. Maeda and Sato
\cite{MS83a,MS83b} investigated the expansion of a shell of zero
thickness and finite surface density of matter inside a spatially homogeneous
dust medium with different densities on each side of the shell. They derived the
equation of motion of the shell and the equation for mass-accumulation in the
shell. They solved these equations numerically for the three types of FLRW
background. The dependence of the enlargement of the void on the time of its
formation was derived. In general, the earlier the formation time, the larger
the enlargement. Moreover, the enlargement is increased for higher background
density.

When studying the light propagation within Swiss Cheese models an important role
is played by the proper randomisations. In some papers, like in the model by
Marra et al., structures are lined up. However, as shown in
\cite{BTT07a,BTT07b,VaFW2008,Szyb2010,Bole2011a} if one allows for randomisation
of structures and angles at which light rays enter the structures, then the
effect of inhomogeneities on the distance-redshift relation is reduced.

An intriguing result was presented in \cite{ClFe2009}, where the Swiss Cheese
model was constructed using the Schwarzschild solution. Their approach is a
generalisation of the Lindquist \& Wheeler model \cite{LiWh1957} and aims at
describing the matter content of the Universe, which in fact is not a continuous
fluid. The result, however, is that the distance to a given redshift in this
model is smaller than in a homogeneous perfect fluid model. Thus in order to fit
the supernova data, more dark energy is needed. For more details see \cite{CC2}.

To escape the spherical symmetry of the vacuoles, a generalisation to the
Szekeres Swiss-cheese models was proposed in \cite{BoCe2010}. As a first step,
and for simplicity, particular classes of axially symmetric quasi-spherical
Szekeres holes were used to reproduce the apparent dimming of the supernovae of
type Ia. The results were compared with those obtained in the corresponding LT
Swiss-cheese models. Although the quantitative picture is different, the
qualitative results are comparable, i.e, one cannot fully explain the dimming of
the supernovae using small scale ($\sim 50$ Mpc) inhomogeneities. To fit
successfully the data, structures of at least $\sim 500$ Mpc size are needed.
However, this result might be an artifact of axial light rays in axially
symmetric models (the model is not fully general). This work is a first step
toward using the Szekeres Swiss-cheese models in cosmology.

\section{The inverse problem}\label{InvPrb}

The inverse problem is conceptually different from the direct approach. Here one
does not parametrise a model and look for the best fit values of assumed
parameters. Instead, one uses observations to specify the functions that define
the model. This idea was pursued by Kristian \& Sachs \cite{KrSa1966}, who were
the first to consider how to use observations to determine the geometry of the
Universe. They used series expansions in powers of the diameter distance and
focused on such observables like redshift, image distortion, number density and
proper motion. The problem was revived by Ellis et al
\cite{EN85,StEN1992,MHMS96,AABFS01,AruSto08,AS09}. They considered the fluid-ray tetrad and
focused on the spherically symmetric case and its perturbations. For the review
and pedagogical presentation of the fluid-ray tetrad problem see
\cite{HeAl2009}.

\subsection{Distance}

The simplest version of the inverse problem is just to take the distance
measurements (angular or luminosity) and use it to define the model. This
approach is mostly based on the LT model. However, to define such model one
needs 2 functions. This means that using just distance measurements one of the
functions needs to be specified by an ansatz instead of by observations. The
simplest ansatz is to assume a spatially flat LT model. An LT model with $E(r) =
0 = \Lambda$ was considered in \cite{Cel00,IgNN2002,VFWa2006}. The model was
fitted to the luminosity distance-redshift relation alone. This implies
constraints on $t_B(r)$ which were given either in terms of constraints on the
lower order derivatives of $t_B(r)$ taken at the observer as in \cite{Cel00} or
in terms of differential equations which were numerically solved as in
\cite{IgNN2002,VFWa2006}. Reference \cite{IgNN2002} also presented an algorithm
defining the LT model from distance measurements and the assumption $t_B =0$.

The aim of this approach was to show that supernova observations alone imply
neither dark energy nor accelerated expansion of the Universe. However, by
imposing additional constraints some tried to argue otherwise. For example if
one imposes smoothness conditions, i.e. density profile at the origin with
vanishing first derivative, then one obtains that the luminosity distance within
the LT model and the FLRW model are the same up to the second order
\cite{TaNa2007}, which implies that the deceleration parameter for pure dust
models must be positive. This, however, does not have any serious cosmological
implications as, firstly, density does not need to be smooth \cite{Roma2010},
and, secondly, models with a smooth density profile can also fit the data
without dark energy (for example most of the giant void models have a smooth
density distribution, see Sec. \ref{giantvoid}).

The inverse problem that uses the angular diameter distances (this relates also
to Sec. \ref{disnz} -- \ref{disnztz}) has difficulties at the apparent horizon
(AH), where $\tdil{\Rh}{z} = 0$. As this quantity can appear in the denominator
(with another quantity vanishing at the AH in the numerator) this can cause
problems when numerically solving the equations. This is not a drawback of the
model, as improperly claimed in \cite{VFWa2006}, and this can be dealt with in
several ways. One of the solutions is to employ the Taylor expansion at the AH
\cite{LuHe2007,McHe2008,BoHA2011}. Another approach is based on solving the equations
on both sides of the AH and choosing such solutions that approach each other
\cite{YoKN2008}. In \cite{CeBK2010} the problem was solved by fitting
polynomials to the functions  $M(r)$ and $E(r)$ and using them as initial
conditions for the direct method.

The existence of the AH can in fact be useful as it puts additional constraints
that must hold at this location. For example for the Lema\^itre model (and its
subcase the LT model) we have \cite{AH09}
\begin{equation}
6 M = 3 R - \Lambda R^3. \label{e33}
\end{equation}
A generic set of data will not obey the above relation. Also, as discussed in
\cite{MHE97,YooC2010}, there are some other relations that will not be satisfied
by generic data because real observational data are always accompanied by
systematics. Thus, these relations can be used to estimate a correction for
systematics so that a consistent solution is obtained. The algorithm for such
corrections is presented and discussed in \cite{LuHe2007,McHe2008,BoHA2011}.

\subsection{Distance and galaxy number counts}\label{disnz}

An algorithm which shows how to define an LT model based on distance and number
count data was first presented in \cite{MHE97}. The algorithm was further
developed in \cite{LuHe2007,McHe2008} but no real observational data has been
used. In \cite{CeBK2010} this algorithm was applied to $D(z)$ and $n(z)$ of the
same form as in the $\Lambda$CDM model. In such a case the model obtained does
not exhibit a giant void. The density
at the current instant in this case is slowly increasing up to $\delta \approx
0.05$ and then is decreasing with an overall profile more resembling a hump than
a void. The bang time function is negative and decreasing to around -2 Gyr at 4
Gpc.

In \cite{KoLa2009}, the same goal of reconstructing an LT model from the
luminosity-distance-redshift relationship and the light-cone matter density as a
function of redshift that matches the fiducial $\Lambda$CDM model was pursued.
The results exactly agree with those of \cite{CeBK2010}. Another result of this
paper is that the LT model whose $D_L(z)$ and $\rho(z)$ functions exactly match
those of the fiducial $\Lambda$CDM model has singular initial conditions for
$R,_r$, which means that $R,_r \rightarrow + \infty$ when the bang time is
approached away from the center, i. e., $R,_r(r\neq0, t_B(r))=+ \infty$.

\subsection{Distance and expansion rate}\label{dishz}

Reference \cite{CeBK2010} also described an algorithm for defining a model based
on distance and expansion rate observations. Again it was assumed that $D(z)$
and $H(z)$ are the same as in the $\Lambda$CDM model. The results suggested a
model with a hump rather than a void, with a  decreasing bang time function.

\subsection{Distance and age of the Universe}\label{distz}

The first attempt to use real data to define the LT model was presented in
\cite{BoHA2011}. Up to date there are no precise measurements of galaxy number
counts, also the measurements of $H(z)$ \cite{SiVJ2005,SJVKS20009} are based on
the assumption that $t_B =0$ and cannot be used to define a general LT model.
Therefore an algorithm for defining an LT model from distance and age
measurements is given in \cite{BoHA2011}. The paper discusses two separate cases
with and without the cosmological constant. In the case of $\Lambda =0$ the
results are somewhere between the giant void and hump configurations, i.e. the
present-day density profile initially increases as in giant void models, but
then decreases. However, due to poor data at high redshift one cannot have
confidence in the model at large distances. The constraints on $t_B$ are not
tight and are consistent with either increasing or decreasing profiles. When
$\Lambda \ne 0$ the results suggest a very slowly increasing profile, but are
consistent with a homogeneous configuration.

\subsection{Distance and redshift drift}\label{disdz}

An algorithm defining an LT model based on distance and redshift drift data
(both as functions of redshift) can be found in \cite{ArSt2010}. The model uses
the fluid-ray tetrad approach \cite{HeAl2009}.

\subsection{Distance, galaxy number counts, and age of the Universe}\label{disnztz}

To specify the LT model one needs to know 2 functions and 1 parameter (the
cosmological constant, which usually, within the LT framework, is set to be
zero). Ref. \cite{BCK2011} presented the algorithm how to specify the LT model
with the cosmological constant based on the distance, galaxy number counts, and
age of the Universe data. Using 3 sets of data allows to break the degeneracy
(described in Secs. \ref{disnz} and \ref{dishz}) between the $\Lambda$CDM model
and zero-$\Lambda$ LT model.

\subsection{Consistency between observations}\label{consch}

Another approach to study observations (instead of directly fitting a model with
them) is based on checking the consistency between observations, i.e. to check
if the relation between observations is as given by the cosmological model.
Ribeiro and Stoeger considered the consistency between the galaxy luminosity
function and corresponding galaxy number counts  \cite{RiSt2003}. In a follow-up
paper they showed that such an analysis strongly depends on the distance
definition used \cite{AIRS2007}. Clarkson, Bassett \& Lu \cite{ClBH2008} studied
the relation between $H(z)$ and $D(z)$ data. They found that if the Universe is
almost homogeneous on large scales, then the expansion rate and distance are not
independent but are related. Thus, by studying relations between observations
one can test the large scale homogeneity of the Universe. An additional problem
arises when the observed objects evolve. A discussion of a possible distinction
between the effect of evolution and inhomogeneity was presented in
\cite{He2001}.

The motivation for the consistency checks is that the relation between different
sets of observational data does not have to be the same as in the cosmological
model that we assume to analyse the data. In \cite{BCK2011} it was shown how
using 3 different sets of data we can test consistency between observations and
the underlying background cosmological models.

\section{What if the cosmological constant is not zero?}

In most of the literature applying inhomogeneous models to fit the observations
the cosmological constant has been set to be zero. Actually, the aim of these
works was to get rid of the impenetrable dark energy component. However, if, for
some theoretical reason, coming for example from particle physics, a nonzero
cosmological constant appeared to be part of the Universe energy budget, the
effect of the inhomogeneities observed in the Universe should still be taken
into account to build a proper cosmological model. Actually, the studies
realised up to now show that their influence is not negligible.

Marra and Paakkonen \cite{MaPa2010} studied the giant void models with a
non-zero cosmological constant. Their conclusion is that if $\Omega_\Lambda
\approx 0.7$ then large voids are excluded by cosmological observations. On the
other hand, large voids ($R_v \sim 3$ Gpc) with $\Omega_\Lambda \approx 0 - 0.3$
are consistent with the data.

Models with Mpc-scale inhomogeneities and cosmological constant were considered
in  \cite{BoCe2010}, where it is shown that smaller values of $\Lambda$ (than
when homogeneity is assumed) are sufficient to fit the data. This is because
small-scale inhomogeneities lead to an increase of the distance (see also Fig.
\ref{discor}) hence less dark energy is needed \cite{BoCe2010}. However, if the
CMB constraints are taken into account, the opposite is true -- in order to have
a good fit more dark energy is needed (than when homogeneity is assumed)
\cite{Bole2011b}. Also, as shown
in \cite{AKMQ2010}, adding inhomogeneities to a model with the cosmological
constant can actually improve the fit to the data, compared to purely
homogeneous models.

The above mentioned studies are based on the direct approach. The first inverse
approach with the pre-assumed cosmological constant was presented in \cite{BoHA2011}.
The full inverse problem that uses the data to derive also the value of the
cosmological constant was discussed in Ref. \cite{BCK2011}.

\section{Formation of black holes}

When studying black holes it is commonly assumed that these objects can be
described using the Schwarzschild or Kerr metrics. This approach has the
following caveats: (1) these space-times are asymptotically flat while the real
Universe is not; (2) these black holes do not evolve, they exist unchanged from
$t = -\infty$ to $t = + \infty$, while real black holes accrete mass.

The solution for the first problem are superpositions of the FLRW models with
stationary black holes such as the Swiss cheese Einstein--Straus \cite{EnSt1945}
configuration. Still, such black holes do not evolve, they exist {\it ab initio}
and their masses do not change, whereas in cosmology we are interested in
evolving black holes and in their formation.

An LT model can solve both these problems. Its first application to a study of
the formation of black holes was presented in \cite{KrHe2004b}, and then
followed by \cite{FR08,Forp2011}. Using it, one can study the evolution of
primordial black holes
or both the formation and evolution. For the most detailed analysis see
\cite{JaKr2011}. The process analysed in detail in Refs. \cite{KrHe2004b} and
\cite{JaKr2011} was predicted by Bondi \cite{Bond1947} already in 1947. A black
hole is formed because rapidly collapsing matter forces the light rays to also
converge toward the final singularity. A black hole with mass comparable to
those at the centres of galaxies may form either out of a localised mass-density
perturbation, or out of a localised velocity perturbation, or around a
pre-existing wormhole \cite{KrHe2004b}. In each case, an apparent horizon is
formed because of the rapid collapse, and the collapse is caused either by
gravitational attraction of the initial condensation, or by the initial
fluctuation of velocity that magnifies itself in the course of collapse.

So far the problem was not considered beyond the LT models. Although the
collapse within the Szekeres model was studied \cite{Szek1975b} it was only
within the asymptotically flat models, not within a cosmological background.

\section{Observational predictions}

There are a number of potentially observable effects that could occur only when
inhomogeneities are present and do not exist in the Friedmann models. The best
known among them is gravitational lensing (see paragraphs 4 and 5 of Sec.
\ref{disfut}). In this section we are not going to discuss the effects that are
most often modelled using perturbative methods, such as gravitational lensing or
the Rees--Sciama effect. Instead, we will focus on the less known effects, in
particular those that can potentially be used to distinguish between Gpc-scale
inhomogeneous models and homogeneous models with dark energy. Thus, the list
below is very selective and does not include all possible observational tests.

\medskip

\noindent $\bullet$ \hspace {3mm} Redshift drift

\noindent As the Universe evolves, the redshifts of astronomical objects change
with time. For the $\Lambda$CDM model $\Delta z >0$ for $z<2$. For the giant
void models (which are the most popular alternative among the inhomogeneous
models) $\Delta z$ is expected to be negative for all $z$. Thus, a detection of
a negative redshift drift for all $z$ would be a proof against dark energy.
However, the converse is not true, as there are Gpc-scale inhomogeneous models
that also have $\Delta z >0$ for low $z$ \cite{YoKN2010}.

\medskip

\noindent $\bullet$ \hspace {3mm} Galaxy number counts

\noindent The galaxy distribution on small scales is very inhomogeneous, with
large fluctuations in number counts. However, with the increasing amount of data
we should be able to detect an overall trend of $n(z)$. In this case it will be
possible to see if the overall behaviour is consistent with the prediction of
homogeneous models. Although a  detection of a Gpc-scale inhomogeneous trend
would be an argument against large-scale homogeneity, the converse argument does
not hold as there are inhomogeneous models that can have the same $n(z)$ as
homogeneous models \cite{CeBK2010}.

\medskip

\noindent $\bullet$ \hspace {3mm} Kinematic Sunyaev-Zel'dovich effect

The existence of a Gpc-scale inhomogeneity leads to an additional (compared to a
homogeneous scenario) peculiar velocity of galaxy clusters. As discussed in Sec.
\ref{giantvoid}, the present data already puts tight constraints on the size of
such an inhomogeneity. Thus, with new data coming from the Planck mission, the
giant void models will be put to the test.

\medskip

\noindent $\bullet$ \hspace {3mm} Ly$\alpha$ observations

\noindent Observations of Ly$\alpha$ lines in spectra of distant quasars provide
information about the amount of light elements. These observations can be used
to constrain cosmological parameters, for example the D/H ratio is very
sensitive to $\Omega_b h^2$. The accurate analysis of the observations is
difficult as the amount of light elements also depends on astrophysical
processes. However, it is believed that low metallicity objects should have
the deuterium to hydrogen ratio unchanged from the time of the primordial
nucleosynthesis.

Within a homogeneous universe $\Omega_bh^2$ should be everywhere the same. The
observations, however, show a large scatter in the data which is also not
consistent with the WMAP data \cite{PZMLS2008}. A conventional explanation of
this phenomenon is that the errors in the individual measurements of D/H may
have been underestimated \cite{Stei2007,PZMLS2008}. As this may be true, in the
future, with a large amount of data and more precise observations, it will be
possible to detect if the variation of $\Omega_bh^2$ is real.

\medskip

\noindent $\bullet$ \hspace {3mm} Dark flow

\noindent In the standard approach, the galaxy velocity field is described using
linear perturbations of the Friedmann model. Within this framework, flows of
large amplitude on a scale beyond 100 Mpc are exceptional. However, observations
show that such a flow exists on scales of at least 150 Mpc
\cite{WaFH2009,FeWH2010}. Although such a flow is hard to explain using linear
methods, it may still be consistent with the standard cosmological model. But if
this flow extends to even larger scales, the $\Lambda$CDM model will not be able
to account for it.

Recently, Kashlinsky {\it et al.} reported the existence of flows on scales over
at least 800 Mpc \cite{KAKE2008,KAEEK2010}. The result of their analysis is
subject to large systematics and so far has not been confirmed by any other
group. However, if such a flow is confirmed, then this will put the $\Lambda$CDM
model at odds with the data.

\medskip

\noindent $\bullet$ \hspace {3mm} Maximum of the diameter distance

\noindent The position of the maximum of the angular distance puts additional
constraints on a model, see (\ref{e33}). This relation combines the distance,
mass and the cosmological constant \cite{Hell2006}. Thus it may serve as a
consistency check and may be used to rule out the models that do not meet this
criterion. Also, the position itself can be different for different types of
models. For example, the giant void models have typically the maximum around $z
\approx 1$ while the $\Lambda$CDM model has a maximum around $z=1.6$
\cite{BoWy2009}.

\medskip

\noindent $\bullet$ \hspace {3mm} Non-repeatable light paths (non-RLPs).

\noindent In  \cite{KrBo2010} it was shown that within inhomogeneous models
generic light rays do not have repeatable paths: two rays sent from the same
source at different times to the same observer pass through different sequences
of intermediate matter particles. This effect does not exist in the
Robertson--Walker models. This shows that RLPs are very special and in the real
Universe should not exist. As a consequence, cosmological objects should change
their positions in the sky. Although the effect is small, in principle it is
detectable.

The existence of this effect may also have consequences in applying averaging
schemes. Within an averaging scheme, an inhomogeneous distribution is
approximated with a uniform (averaged) model. As a fist approximation it is
assumed that light propagates along null geodesics of a homogeneous model (the
only difference is that the evolution of the model is governed by the Buchert
rather than Friedmann equations). However, if geodesics that join the observer
and the source proceed at different times through different sequences of
intermediate matter particles, then the path of the light ray within an average
geometry may not be a geodesic anymore.

\section{Pervasive errors and misconceptions}\label{errandmisc}

Many astrophysicists tolerate a loose approach to mathematics and physics.
Papers written in such a style planted errors and misconceptions in the
literature, which were then uncritically cited in other papers and came to be
taken as established facts. In this section we present a few most damaging
misconceptions (marked by black squares $\blacksquare$) together with their
explanations (marked by large asterisks {\huge \bf $_*$}).

\medskip

\noindent $\blacksquare$ \hspace {3mm} The LT models that explain away dark
energy with matter inhomogeneities contain a ``weak singularity'' at the centre
\cite{VFWa2006}, where the scalar curvature $R$ has the property $g^{\mu \nu}
R;_{\mu \nu} \to \infty$.

\medskip

\noindent {\huge \bf $_*$} \hspace {3mm} $g^{\mu \nu} R;_{\mu \nu} \to \infty$
is not a singularity by any accepted criterion in general relativity
\cite{KHCB2010}. It only implies a discontinuity in the derivative of mass
density by distance -- a thing quite common in Nature (e.g. on the surface of
the Earth). At the centre, $g^{\mu \nu} R;_{\mu \nu} \to \infty$ implies a
conical profile of density -- also a nonsingular configuration.

\medskip

\noindent $\blacksquare$ \hspace {3mm} Decelerating inhomogeneous models with
$\Lambda = 0$ cannot be fitted to the same distance--redshift relation that
implies acceleration in $\Lambda$CDM. This is because a certain equation
connecting the deceleration parameter $q_4$ to density, expansion and shear
prohibits $q_4 < 0$ \cite{HiSe2005}.

\medskip

\noindent {\huge \bf $_*$} \hspace {3mm} The equation derived in \cite{HiSe2005}
(formally analogous, but inequivalent, to the Raychaudhuri equation) is based on
approximations that are not explicitly spelled out \cite{KHCB2010}. An
approximate equation cannot determine the sign of anything. If the
approximations are taken as exact constraints imposed on the LT model, they
imply zero mass density, i.e. the Schwarzschild limit. Moreover, the $q_4$ of
\cite{HiSe2005}, although it coincides with the deceleration parameter in the
Friedmann limit, is not a measure of deceleration in an inhomogeneous model (it
is defined by the Taylor expansion of the luminosity--redshift relation). Refs.
\cite{BoHA2011,CeBK2010,IgNN2002} provide an explicit demonstration that a
decelerating LT model with $\Lambda = 0$ {\it can} be fitted to {\it exactly the
same} distance - redshift relation that holds in the $\Lambda$CDM model. This
relation is reproduced by a spatially inhomogeneous expansion pattern, without
any dark energy.

\medskip

\noindent $\blacksquare$ \hspace {3mm} There is a ``pathology'' in the LT models
that causes the redshift-space mass density to become infinite at a certain
location (called ``critical point'') along the past light cone of the central
observer \cite{VFWa2006}.

\medskip

\noindent {\huge \bf $_*$} \hspace {3mm} The ``critical point'' is the apparent
horizon (AH), at which the past light cone of the central observer begins to
re-converge toward the past. This re-convergence had long been known in the FLRW
models \cite{Elli1971,McCr1934}, and the infinity in density is a purely
numerical artifact -- a consequence of trying to integrate past AH an expression
that becomes 0/0 at the AH. Ways to handle this problem are known
\cite{LuHe2007,McHe2008, CeBK2010}.

\medskip

\noindent $\blacksquare$ \hspace {3mm} Fitting the LT model to cosmological
observations, such as number counts or the Hubble function along the past light
cone, results in predicting a huge void, at least several hundred Mpc in radius,
around the centre (see discussion in Sec. \ref{giantvoid}).

\medskip

\noindent {\huge \bf $_*$} \hspace {3mm} The implied huge void is a consequence
of handpicked constraints imposed on the arbitrary functions of the LT model,
for example a constant bang time $t_B$. With no a priori constraints, the giant
void is not implied \cite{CeBK2010}.

\medskip

\noindent $\blacksquare$ \hspace {3mm} The bang time function must be constant
because $\dril {t_B} r \neq 0$ generates decaying inhomogeneities, which would
have to be ``huge'' in the past, and this would contradict the predictions of
the inflationary models (private communication from the referees of
\cite{CeBK2010}).

\noindent {\huge \bf $_*$} \hspace {3mm} While it is true that in models with
$\dril {t_B} r \ne 0$ the early Universe was very inhomogeneous, it does not
mean that such models could not be realistic (so far they are consistent with
observations after all). Although in the current paradigm the early Universe
undergoes inflation that is supposed to leave it very homogeneous, the
occurrence of inflation is not in any way proven. Inflationary models are just
one of hypotheses that compete for observational confirmation. Thus, using them
to justify or reject some other hypotheses may sound dogmatic and is in fact
un-scientific.

\section{Discussion and future prospects }\label{disfut}

We have seen that one can find in the literature a number of models constructed
with exact inhomogeneous solutions of Einstein's equations which
fit the available observational data as properly as (and sometimes better than)
the standard $\Lambda$CDM model.

The LT model with a central observer, which is sometimes criticised as being at
odds with the Copernican Principle, must be, in our view, only considered as an
intermediate model where the angular inhomogeneities have been smoothed around
the observer and only the radial inhomogeneities have been taken into account
(an example of such a situation is presented and discussed in \cite{BoSu2010}).
Moreover, the use of oversimplified LT models can create another false idea and
false expectation. The false idea is that there is an opposition between the
$\Lambda$CDM model, belonging to the FLRW class, and the LT model or in general,
inhomogeneous models: it is believed that either one or the other could be
`correct', but not both. This putative opposition can then give rise to the
expectation that more, and more detailed, observations will be able to tell us
which one to reject. In truth, there is no opposition. The inhomogeneous models,
like for example the LT model with its two arbitrary functions of one variable,
are huge, compared to FLRW, {\em families} of models that include the Friedmann
models as a very simple subcase. The fact, demonstrated in several papers, that
even a $\Lambda = 0$ LT model can mimic $\Lambda \neq 0$ in an FLRW model,
additionally attests to the flexibility and power of the LT model. Thus, if the
Friedmann models, $\Lambda$CDM among them, are considered good enough for
cosmology, then the LT models can only be better: they constitute an {\em exact
perturbation} of the Friedmann background, and can reproduce the latter as a
limit with an arbitrary precision. The right question to ask is not ``which
model to reject: FLRW or LT?'', but ``how close to their FLRW limits must the LT
arbitrary functions be to satisfy the observational constraints?''.

Nature does not create objects that fulfil mathematical assumptions with perfect
precision. Objects in mechanics or electrodynamics that are described as
spherically symmetric have this symmetry only up to some degree of
approximation. An ``ideal gas'' in thermodynamics is nearly ideal only at
sufficiently low pressure. An ``incompressible fluid'' ..., and so on. Why
should the Universe be an exception and be exactly homogeneous in the large (and
exactly spatially flat in addition)?

In fact, we already have qualitative evidence that our observed Universe is not
FLRW: the gravitational lenses. The FLRW models are conformally flat, so the
null geodesics in them are conformal images of the null geodesics from the
Minkowski spacetime. In this spacetime, rays sent from a common origin never
intersect again. So, in a conformally flat spacetime rays issuing from a common
source can intersect again only in such points that are singularities of the
conformal mapping (and, consequently, of the spacetime itself). Then, however,
the positions of the points of second intersection are determined by the
geometry of spacetime, and not just by the initial points and directions of the
rays, as is the case in a gravitational lens (where, in addition, there is no
spacetime singularity at the intersection point). Hence, a spacetime containing
a gravitational lens cannot be conformally flat.

Gravitational lenses are observed in our Universe at the distance scales, at
which the FLRW approximation is supposed to already apply, namely the lensing
objects and the sources of lensed rays are quasars. So, our Universe does not
have the FLRW geometry at large scales.

One more qualitative evidence of our Universe being non-FLRW on large scales may
be provided by the effect of non-repeatable light paths, described in Ref.
\cite{KrBo2010}.

It is strange that a large part of the astrophysical community is comfortable
with the idea of linearised perturbations around homogeneous models, but reacts
with strong negative emotions to exact perturbations represented by
inhomogeneous models.

In the future, the LT models will be used to extract the cosmic metric from
observations. This programme has been initiated in
\cite{LuHe2007,McHe2008,BCK2011}. This is the full inverse problem. To date it
has been assumed that the metric has the LT form, as a relatively simple case to
start from, but the long term intention is to remove the approximation of
spatial symmetry.

All known exact solutions of the Einstein equations which can be of cosmological
use possess some symmetries or quasisymmetries. The only way to overcome such
shortcomings is to obtain a fully operational, exact and inhomogeneous solution
of these equations. This can only be achieved using numerical relativity and we
suspect that this will be the new way of dealing with cosmology in the years to
come.

\ack This research was partly supported by the Marie Curie Fellowship under the
grant PIEF-GA-2009-252950 (KB) and by the Polish Ministry of Higher Education
grant N N202 104 838 (AK).

\end{document}